\documentclass[%
aps,
 reprint,
 amsmath,amssymb,
prper,
floatfix
]{revtex4-2}

\usepackage{graphicx}
\usepackage{dcolumn}
\usepackage{bm}
\usepackage{tikz}
\usepackage{pgfplots}
\usepackage[export]{adjustbox}
\pgfplotsset{compat=1.17}
\usepackage{hyperref}

\begin{document}


\title{Ethel: A Virtual Teaching Assistant}

\author{Gerd Kortemeyer}
 \email{kgerd@ethz.ch}
 \affiliation{%
Rectorate and AI Center, ETH Zurich, 8092 Zurich, Switzerland
}%
\altaffiliation[also at ]{Michigan State University, East Lansing, MI 48823, USA}
\date{\today}

\begin{abstract}
Note: The following article has been submitted to The Physics Teacher. After it is published, it will be found at \url{https://pubs.aip.org/aapt/pte}
 \end{abstract}
\maketitle

Generative AI has shown potential in solving physics problems~\cite{kortemeyer23ai,yeadon2024impact} and providing feedback on assessments~\cite{kortemeyer24aigrading}. However, results are sometimes still inaccurate~\cite{kuchemann2023can}, at the wrong level, or using notations and definitions not appropriate for a particular course. A possible solution is augmenting the prompts with course-specific reference materials. Also, for feedback on homework solutions and grading exams, the problem text and the sample solution or grading rubric can be injected into the prompts~\cite{kuchemann2024large}. Project Ethel at ETH Zurich aims to construct a virtual teaching assistant using these practices.

\section{Chatbot}
Ethel's chatbot utilizes course-specific materials (typically PDFs) through Retrieval Augmented Generation (RAG)~\cite{lewis2020retrieval}, employing tools such as LangChain~\cite{langchain} and OpenAI's ada embeddings~\cite{openaiada}; this can be implemented in about 400 lines of code~\cite{ethelrag,ethelblog}. Instead of complex tuning, these materials are stored in a reference database accessible to the bot, which uses a Large Language Model (LLM) for interaction.

In multilingual contexts like Switzerland, proficiency in several languages is essential, hence the use of OpenAI's commercial GPT-4~\cite{gpt4}. For monolingual English applications, similar RAG functionality is achievable with open-weight models like Llama~\cite{llama}. Regardless of the model type, running any LLM (``inference'') round-the-clock and at-scale currently requires commercial cloud-based platforms; while university high-performance computer clusters might have the necessary GPU-power, they are usually designed for batch operation. This raises data privacy concerns, currently addressed through the university's agreement with Azure AI Services~\cite{azure}.

\begin{figure*}
\includegraphics[width=\textwidth]{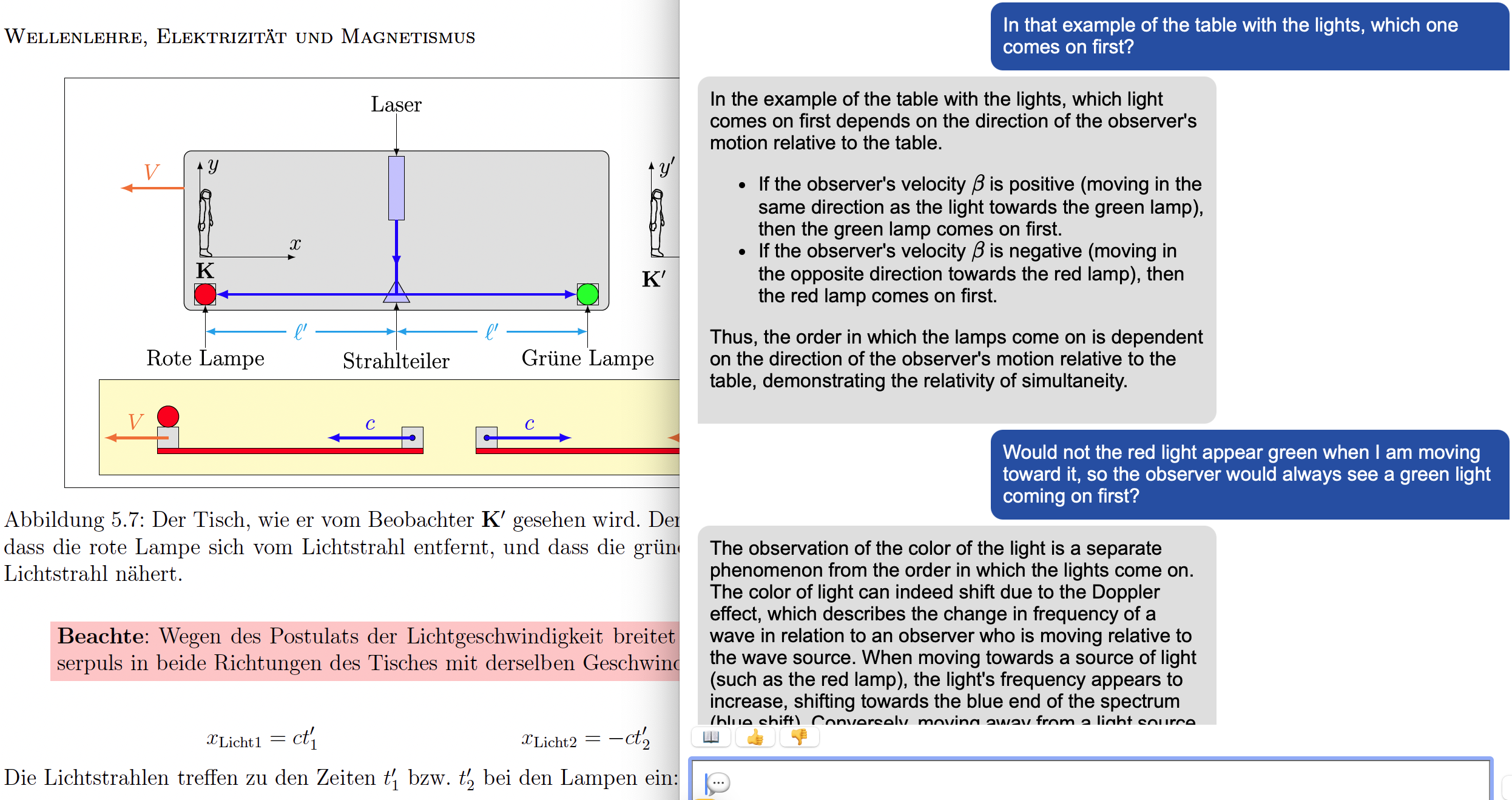}
\caption{Example of the chatbot component of Ethel. Shown on the left is a short excerpt of the lecture script, on the right a dialogue with the system.}
\label{fig:chatbot}
\end{figure*}

Figure~\ref{fig:chatbot} depicts a dialogue where the prompt inquires about ``that example of the table with the lights.'' This prompt is adequate for the bot to pinpoint the relevant sections from the course's 300-page German-language script, displayed on the left side. Despite a subsequent attempt to confuse the bot involving the Doppler shift, it accurately identified the reference and returned to the original topic.

Ethel was deployed for over 1500 students across six courses during the Spring semester of 2024, including two large introductory physics courses. Although the system sometimes provided incorrect answers, a survey revealed that the majority of students still found it helpful or very helpful. Despite concerns about misplaced trust in bots like Ethel~\cite{dahlkemper23,ding2023students}, an earlier survey among 4800 of our students indicated they were well aware of the limitations of LLMs and maintained healthy skepticism towards the responses~\cite{balabdaoui2024survey}.
A critique emerged that RAG was almost too efficient for study purposes, as it quickly provided specific definitions and statements from the lecture materials, exactly what students needed assistance understanding in the first place.

\section{Homework Feedback}
\begin{figure*}
\includegraphics[width=0.76\textwidth]{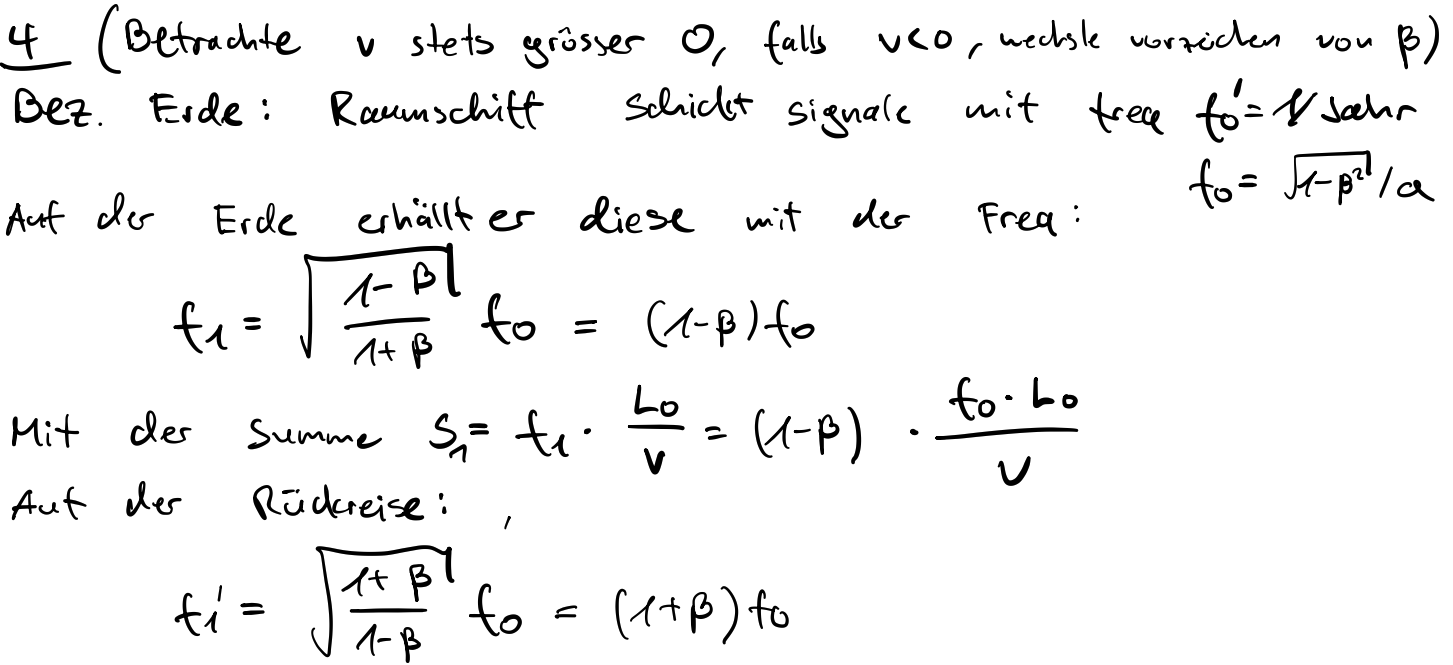}
\caption{Student solution to a homework problem.}
\label{fig:student}
\end{figure*}

Ethel's homework component gives students feedback on handwritten  solutions like the one shown in Fig.~\ref{fig:student}.  Students could upload these as scanned PDFs to a course management system. For our experiment, we converted them to LaTeX using a combination of Mathpix~\cite{mathpix} and GPT-4V~\cite{gpt4v}; in the future, we will use GPT-4o~\cite{gpt4o}.

\begin{figure*}
\includegraphics[width=0.6\textwidth]{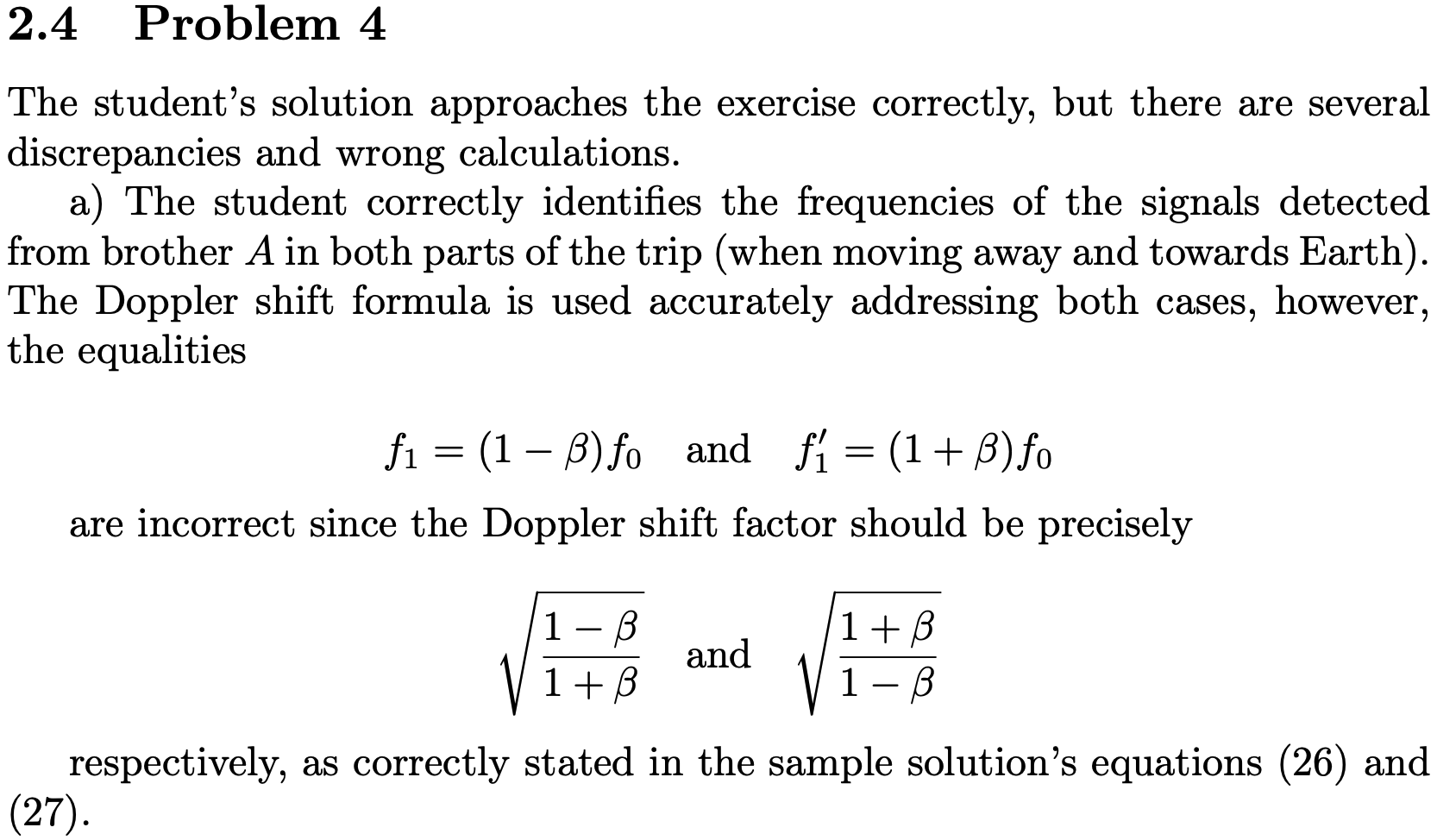}
\caption{Ethel feedback to the student solution in Fig.~\ref{fig:student}, based on and referencing the sample solution.}
\label{fig:feedback}
\end{figure*}

Based on the sample solution provided by the instructor, the system then produces feedback like the one shown in Fig.~\ref{fig:feedback}. Initially, the system was prompted to address the student directly and be ``encouraging,'' but over future iterations, prompts were increasingly rewritten to provide more impersonal feedback. The novelty effect of anthropomorphizing the system quickly wore off, and the system also had a tendency to use patronizing formulations and give unsolicited study advice.

Students rated the feedback to be helpful  and correct in 3/4th of the cases. Ethel tended to underestimate the correctness of the students' solutions, where the issue was mostly handwriting recognition, which students only rated as accurate in about half of the cases.

\section{Exam Grading}
Another task of TAs is exam grading. In a study involving the grading of a high-stakes thermodynamics exam for $252$ students, we garnered insights on that~\cite{kortemeyer2024grading}. The study highlighted that while LLMs struggle with numerical tasks such as counting and addition, they are beneficial for grading free-form exam responses due to their ability to handle fuzzy data probabilistically. Also here, though, the primary challenge was handwriting recognition.

Further analysis revealed that grading prompt granularity significantly impacts the success of automated grading. Using a fine-grained rubric for entire problems often resulted in bookkeeping mistakes and grading failures. In contrast, dividing problems into several parts and using a comprehensive sample solution proved more reliable, although it occasionally missed nuanced details and specific rubric weightings. Grading graphical solutions, like process diagrams, was substantially less reliable than grading mathematical derivations, with inaccuracies often due to extraneous visual elements in freehand drawings. Overall, the grading system showed high precision in identifying passing exam solutions but had low recall, missing about half of the other passing solutions.

\section{Outlook}
Operating Ethel cost the university \$7.50 per student per course per semester for Azure AI Services; we will continue to explore open-weight options and possibilities to run services on-premises.
For the chatbot, we plan on implementing an instructor self-serve platform, enabling automatic integration of uploaded documents with the course-specific chatbot. This upgrade is expected to include connectivity with other campus systems, such as lecture recordings, to utilize already generated subtitles seamlessly. Additionally, we plan to introduce direct links to the referenced material segments, allowing students to easily trace the sources of the chatbot's responses. For the assessment components, handwriting and graphics recognition remain the largest challenges.

\bibliography{ethel}
\end{document}